\documentclass[aps,showpacs,floats,amsmath,amssymb]{revtex4}
\usepackage{graphicx}
\usepackage[next]{inputenc}
\usepackage{epsfig}
\usepackage{bm}
\def\be{\begin{equation}}
\def\ee{\end{equation}}

\def\bi{\begin{itemize}}
\def\ei{\end{itemize}}
\def\bn{\begin{enumerate}}
\def\en{\end{enumerate}}
\def\bea{\begin{eqnarray}}
\def\eea{\end{eqnarray}}
\def\no{\nonumber}
\def\ba{\begin{array}}
\def\ea{\end{array}}
\def\bd{\begin{displaymath}}
\def\ed{\end{displaymath}}
\def\la{\langle}
\def\ra{\rangle}

\def\HH{\mathcal{H}}

\def\bs{\bar{s}}
\def\bt{\bar{t}}
\def\bu{\bar{u}}
\def\bd{\bar{d}}
\def\bth{\bar{t}_0}
\def\btau{\bar{\tau}_0}

\def\ua{\uparrow}
\def\da{\downarrow}

\newcommand{\boldtau}{{\bm \tau}}
\newcommand{\bS}{{\bf S}}

\begin{document}
\title{Field induced quantum phase transition in the anisotropic Kondo necklace model}

\author{ P. Thalmeier$^1$ and A. Langari$^{2}$}

\affiliation{
$^1$Max Planck Institute for Chemical Physics of Solids, 01187 Dresden, Germany\\
$^2$Physics Department, Sharif University of Technology, Tehran 11365-9161, Iran}
\date{\today}
\begin{abstract}
The anisotropic Kondo necklace model in 2D and 3D is treated as a genuine model for magnetic to Kondo singlet quantum phase transitions in the heavy fermion (HF) compounds. The variation of the quantum critical point (QCP) with anisotropy parameters has been investigated previously in the zero field case \cite{Langari06}. Here we extend the treatment to finite fields using a generalised bond operator representation including all triplet states. The variation of critical t$_c$ with magnetic field and the associated phase diagram is derived. The influence of anisotropies and the different g-factors for localised and itinerant spins on the field dependence of t$_c$ is also investigated. It is found that three different types of  behaviour may appear: (i) Destruction of antiferromangetism and appearance of a singlet state above a critical field. (ii) The inverse behaviour, namely field induced antiferromagnetism out of the Kondo singlet phase. (iii) Reentrance behaviour of the Kondo singlet phase as function of field strength.
\leftskip 2cm \rightskip 2cm
\end{abstract}
\maketitle

\section{Introduction}
\label{sec1}
The Kondo necklace (KN) type models are useful to discuss the quantum phase transitions between
Kondo singlet and antiferromagnetically ordered states such as found in heavy fermion compounds \cite{Continentino05,Continentinobook,Sachdevbook,Vojta03,Fulde06}. They have  been originally proposed by Doniach \cite{Doniach77} for the one-dimensional case as a simplified version of the itinerant Kondo lattice (KL) models \cite{Tsunetsugu97}. Thereby the kinetic energy of conduction electrons is replaced by an intersite exchange term. For a pure xy-type intersite exchange this may be obtained by a Jordan-Wigner transformation. However in higher D the replacement cannot be justified easily. The intuitive argument is that at low temperatures the charge fluctuations in the Kondo lattice model are frozen out and the remaining spin fluctuation spectrum can be simulated by an antiferromagnetic inter-site interaction term of immobile $\boldtau$ spins coupled by a Kondo interaction to the local noninteracting spins {\bf S}. Recent exact diagonalisation studies on finite clusters for
  both Kondo lattice  and Kondo necklace models have indeed found that the competition between on-site Kondo singlet formation and AF inter-site correlations are very similar in both models \cite{Zerec05,Zerec06}. A more formal way to get rid of charge fluctuations  in the KL model is an inclusion of a Hubbard term for conduction electrons (KLU model) which leads to the isotropic KN model in the large U limit at half filling.

Nevertheless one should consider the Kondo necklace model for D$\geqslant$2 
as a model in its own right which is suitable for studying quantum phase transitions between a Kondo singlet (KS) and an antiferromagnetic (AF) phase. In its original form the local Kondo exchange is isotropic while the intersite exchange is of xy-type. This model has U(1) symmetry. 
Later more general models with arbitrary anisotropy of Kondo as well as intersite exchange terms have been considered \cite{Yamamoto02}. Indeed compounds which exhibit the KS to AF quantum phase transition have mostly uniaxial symmetries. A full account of the influence of uniaxial anisotropies of both terms in the Kondo necklace Hamiltonian on the quantum critical point has been given in Ref. \onlinecite{Langari06}. The D=2 KN model without any anisotropy  may also be understood as a special case of a bilayer Heisenberg model \cite{Kotov98,Bruenger06} where the intersite bonds are cut in one of the layers. A reintroduction of holes in this case leads to the KNtJ model which is related to the KLU model away from half filling \cite{Schork99,Bruenger06}. 

In the general anisotropic KN model the quantum phase transition is achieved by varying the ratio of 'hopping' t, i.e., the intersite interactions of $\boldtau$ spins to the on-site Kondo interaction J. In practice this is achieved by varying pressure (hydrostatic or chemical). An alternative way to arrive at the QCP is to apply an external magnetic field which breaks the local Kondo singlets and leads to a field dependence of  the critical t$_c$. Starting from a noncritical or above critical t at zero field the system may then be tuned to the QCP by varying the field strength. This is indeed a practical method frequently applied \cite{Gegenwart02}.
To investigate field-induced quantum phase transitions from Kondo singlet to AF ordered state or vice versa we
have extended our previous work to include the effect of the magnetic field within the Kondo necklace model. However, this introduces an additional parameter, namely the ratio of effective g- factors for the local Kondo spins and the interacting spins. They can be different due to the different strength of spin-orbit coupling and crystalline electric field effects involved in the formation of the $\boldtau$ and {\bf S} (pseudo-) spins. A mainly numerical study of the itinerant isotropic KL model in a magnetic field has been given previously \cite{Beach04}.

In Sec.~\ref{sec2} we define the anisotropic KN model in a magnetic field and in  Sec.~\ref{sec2a} briefly discuss the local state space of the Kondo term in an external field. In  Sec.~\ref{sec3} we perform the transformation from spin to bosonic variables and in  Sec.~\ref{sec4} derive the selfconsistent equations for the mean field boson condensate amplitudes in the singlet and antiferromagnetic phases.. The influence of higher order terms in the Hamiltonian is discussed in Sec.~\ref{sec4a}. In  Sec.~\ref{sec5} we investigate the numerical solutions and discuss the resulting quantum critical t-h phase diagram. A discussion is provided in Sec.~\ref{sec5a}  and Sec.~\ref{sec6} finally gives the conclusions.


\section{Anisotropic Kondo necklace model in  an external field}
\label{sec2}

To investigate field induced quantum critical behaviour we start from
the anisotropic Kondo necklace (KN) model with U(1) symmetry
\cite{Langari06} where the field is applied along the anisotropy (z)
direction:
\bea
\HH&=&\HH_t+\HH_J+\HH_Z\no\\
&=&t\sum_{<n,m>} (\tau_n^x \tau_m^x + 
\tau_n^y \tau_m^y + \delta \tau_n^z \tau_m^z)
+J\sum_{n}(\tau_n^x S_n^x+\tau_n^y S_n^y + 
\Delta \tau_n^z S_n^z) 
 +\gamma\sum_iS_n^z+\gamma'\sum_i\tau_n^z ,
\label{HAM}
\eea
Here, the summation over nearest neighbor (n.n.) is indicated by brackets and
 $\tau_n^{\alpha}$ is the $\alpha$-component ($\alpha$ = x,y,z) of the 'itinerant'
electron spin at site $n$ whereas $S_n^{\alpha}$ is the $\alpha$-component
of localized spins at position $n$. For the exchange coupling between
the itinerant and localized spins we generally use $J_x\equiv J$ as reference
energy scale in all figures except when stated otherwise.  The local anisotropy parameter 
$\Delta$ is defined by the relation J$_z$=$\Delta$J$_x$ between the z-axis and in-plane (xy) local exchange.  The hopping parameter of the itinerant
electrons is proportional to $t$ with the anisotropy in the
z-direction given by $\delta$. The present model has three control
parameters: $t/J_x$ and the anisotropy parameters
($\delta$,$\Delta$). In the Zeeman term we defined $\gamma=-g_sh$ and 
$\gamma'=-g_\tau h$ with h = $\mu_B$H where H is the strength of the applied field. Furthermore $g_s$ and g$_\tau$ are the gyromagnetic ratios for localised ($S_n^{\alpha}$) and itinerant
($\tau_n^{\alpha}$) spins respectively. They are determined by the
combined effect of spin-orbit coupling and crystalline electric fields
which depends on the degree of localisation or itineracy of
electrons. Therefore g$_s$ and g$_\tau$ in general need not be equal.
We shall consider two extreme cases, namely  ($g_s, g_\tau$)=(2,0) and  ($g_s, g_\tau$)=(2,2).
The former is more realistic, since in real heavy fermion compounds most of the magnetic response is due to
the localised electrons with pseudo-spin {\bf S}.
As in previous work \cite{Zhang00,Langari06} the present study of field
induced quantum phase transitions in the anisotropic KN model is based on the
bond operator formulation. Its Hilbert space is spanned by local
singlet-triplet states of ($\bf S_n,\boldtau_n$) spin dimers represented by
bosonic degrees of freedom. In applying this technique to the finite field case
we will largely use the same or similar notations as in the previous zero field
case \cite{Langari06} for consistency.
%
\section{Local levels and Zeeman splitting}
\label{sec2a}

Before we perform the transformation to bosonic variables it is useful
to have a clear understanding of the local singlet-'triplet' level
scheme as function of anisotropy and magnetic field because certain
tendencies of the quantum critical behaviour are correlated with the
splitting of the ground state and the first excited state of a local bond. Therefore
we first diagonalise the local Hamiltonian $\HH_L=\HH_J+\HH_Z$.  This
leads to eigenvalues $\epsilon_i$=E$_i$/$(J_x/4)$ (i=1-4) given by 
\bea
\epsilon_{1,2}&=&\Delta\pm\hat{\gamma}_+ ,\no\\
\epsilon_{3,4}&=&\pm(4+\hat{\gamma}_-^2)^\frac{1}{2}-\Delta ,\no\\
\hat{\gamma}_\pm&=&-\frac{h}{(J_x/4)}\cdot\frac{1}{2}(g_s\pm g_\tau) 
\equiv\frac{\gamma_\pm}{(J_x/4)} ,
\label{LLEV}
\eea 
with $\gamma_\pm= \frac{1}{2}(\gamma\pm\gamma')=-g_\pm h$ and $g_\pm=\frac{1}{2}(g_s\pm g_\tau)$.
Defining S$^t_z$=$\tau_z$ +S$_z$ then $\epsilon_{1,2}$ are the energies of the triplet states
$|\ua\ua\rangle, |\da\da\rangle$ with S$^t_z=\pm$ 1 respectively which
exhibit the linear Zeeman effect. Furthermore $\epsilon_{3,4}$ correspond to the triplet $S^t_z= 0$ state
$\frac{1}{\sqrt{2}}(|\ua\da\rangle +|\da\ua\rangle)$ and singlet state
$\frac{1}{\sqrt{2}}(|\ua\da\rangle -|\da\ua\rangle)$ respectively. They show level
repulsion in a magnetic field, except for g$_s$=g$_\tau$ when the
total S$^t_z$-component is conserved and commutes with $\HH_L$.
It is obvious that the spectrum of eigenstates in Eq.(\ref{LLEV}) is invariant under the 
transformation (g$_s$,g$_\tau$) $\rightarrow$ (g$_\tau$,g$_s$).
%
\begin{figure}[tbc]
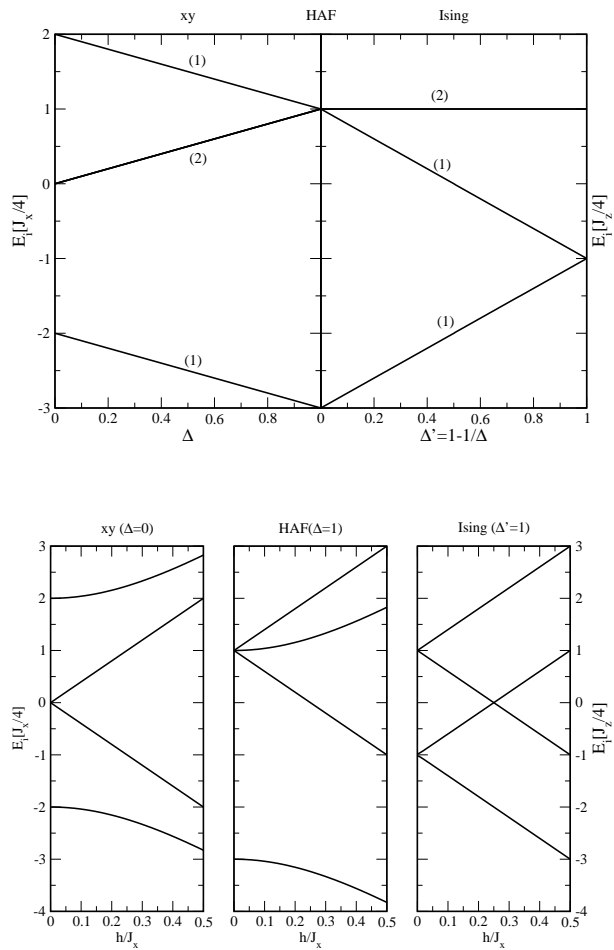

\begin{center}
\includegraphics[width=8.0cm,angle=0]{Fig1a}\\
\vspace{1cm}
\includegraphics[width=8.0cm,angle=0]{Fig1b}
\caption{Top: Dependence of  local energy levels after Eq.~(\ref{LLEV} ) on the local anisotropy
 $\Delta$ (xy-side) or  $\Delta'$=1-1/$\Delta$ (Ising side) at zero field. Numbers in parentheses denote the degeneracy of each level.
Bottom: Field dependence (h=$\mu_B$H) of local energy level for three extreme cases. Note that in the xy-case the excitation energy from ground state to first excited state vanishes asymptotically (h$\gg$J$_x$) whereas in the other cases it reaches a constant (HAF) or is equal to a constant (Ising). Here we used (g$_s$,g$_\tau$)=(2,0).}
\label{Fig1}
\end{center}
\end{figure}
%
The variation of zero-field energy levels with $\Delta$ is shown in
Fig.~\ref{Fig1} (top) from the xy-U(1) limit ($\Delta$=0) via the
Heisenberg SU(2) point($\Delta$=1) to the Ising Z$_2$ limit
($\Delta$=$\infty$). The field dependence for a few selected
$\Delta$-values and g-factors ( g$_s$,g$_\tau$)=(2,0) are
shown in Fig.~\ref{Fig1} (bottom). For  ( g$_s$,g$_\tau$)=(2,2) (not shown in Fig.~\ref{Fig1}) 
the ground state and first excited state  levels cross at h$_{cr}$/J$_x$=0.25, 0.50 for ($\delta,\Delta$)=(0,0) 
and (1,1) respectively. In such a case one may expect that the
quantum critical t$_c$(h$_{cr}$) for the transition from Kondo singlet to AF phase vanishes.

\section{Derivation of effective bosonic model}
\label{sec3}

We apply the transformation from spin variables ($S_\alpha$,$\tau_\alpha$) to 
bond variables (s,t$_\alpha$) ($\alpha$ =x,y,z) to the Hamiltonian in Eq.~(\ref{HAM}).
It is given by \cite{Sachdev90}:
\bea
S_{n, \alpha}=\frac{1}{2}(s^{\dagger}_n t_{n, \alpha}+
t_{n, \alpha}^{\dagger} s_n -i \epsilon_{\alpha \beta \gamma} 
t_{n, \beta}^{\dagger}t_{n, \gamma}), \nonumber \\
\tau_{n, \alpha}=\frac{1}{2}(-s^{\dagger}_n t_{n, \alpha}-
t_{n, \alpha}^{\dagger} s_n -i \epsilon_{\alpha \beta \gamma} 
t_{n, \beta}^{\dagger}t_{n, \gamma}),
\label{BOND}
\eea
where $\epsilon_{\alpha\beta\gamma}$ is the fully antisymmetric tensor. 
By construction the singlet (s) and triplet ($t_\alpha$) operators generate the local eigenstates of $\HH_L$ for zero field and no anisotropy \cite{Yamamoto03} :
\bea
s^\dagger|0\ra&=&\frac{1}{\sqrt{2}}(|\ua\da\ra-|\da\ua\ra);\qquad
t_x^\dagger|0\ra=-\frac{1}{\sqrt{2}}(|\ua\ua\ra-|\da\da\ra),\no\\
t_z^\dagger|0\ra&=&\frac{1}{\sqrt{2}}(|\ua\da\ra+|\da\ua\ra);\qquad
t_y^\dagger|0\ra=\frac{i}{\sqrt{2}}(|\ua\ua\ra+|\da\da\ra).
\eea
Here the first and second arrows indicate the z-component of $\boldtau$ and $\bS$ spins respectively.
The singlet and triplet operators  satisfy the usual bosonic commutation relations according to
\bea
 [s_n, s_n^{\dagger}]=1;\qquad 
 [t_{n, \alpha}, t_{n, \beta}^{\dagger}]=\delta_{\alpha, \beta};\qquad
[s_n, t_{n, \beta}^{\dagger}]=0. 
\eea
All other commutators vanish. The physical states have to satisfy 
the local constraint $s^{\dagger}_n s_n+\sum_{\alpha} t_{n, \alpha}^{\dagger} t_{n, \alpha}=1$. This transformation leads
to an effective Hamiltonian in terms of singlet (s) and triplet (t$_\alpha$) bosons. It
has been argued in \cite{Zhang00,Langari06} that one may restrict to
the terms which are bilinear in the triplet bosons and we will later in Sect.~\ref{sec4a} discuss to which extent this is justified. For the moment we restrict to the bilinear contribution. It can be written as the sum of a local term $\HH_L$ and an inter-site interaction term $\HH_1$. Here
\bea
\HH_L=\HH_J+\HH_Z&=&
\frac{J_x}{4}\sum_n \Big( -(2+\Delta)s_n^{\dagger}s_n +
(2-\Delta)t_{n,z}^{\dagger}t_{n,z}
+\Delta(t_{n,x}^{\dagger}t_{n,x}+t_{n,y}^{\dagger}t_{n,y})\Big)\no\\
&&+\gamma_-\sum_n(s^\dagger_nt_{nz}+t^\dagger_{nz}s_{n})
-i\gamma_+\sum_n(t^\dagger_{nx}t_{ny}-t^\dagger_{ny}t_{nx}) ,
\label{HAML}
\eea
contains the on-site Kondo interaction ($\HH_J$) and the Zeeman term
($\HH_Z$). The bilinear interaction term is given by
\bea
\HH^{(2)}_1&=&\frac{t}{4}\sum_{<n, m>}\sum_{ \alpha=x, y} \Big(s_n^{\dagger} t_{n, \alpha}
(s_m^{\dagger} t_{m, \alpha}+t_{m, \alpha}^{\dagger}s_m)+ h.c.\Big)\no\\
&&+\frac{t \delta}{4}\sum_{<n, m>}\Big(s_n^{\dagger} t_{n, z}
(s_m^{\dagger} t_{m, z}+t_{m, z}^{\dagger}s_m)+ h.c.\Big) .
\label{HAM1}
\eea
The physical constraint on the local Hilbert space is enforced by adding a
Lagrange term at each site with an associated chemical potential
$\mu_n$, leading to
\bea
\HH=\HH_L+\sum_n \mu_n (s_n^{\dagger}s_n+
\sum_{\alpha=x,y,z} t_{n, \alpha}^\dagger t_{n, \alpha}-1)+\HH^{(2)}_1
\equiv \HH_0+\HH^{(2)}_1 .
\label{HAMMU}
\eea
This Hamiltonian is diagonalised within a mean field approximation for
the bond-operator singlet and triplet bosons. We assume that in
general there are three independent bosonic amplitudes which
characterise the phases in the (t,h)-plane: Singlet
$\bar{s}=\la s_n\ra$ denoting the strength of local singlet
formation, staggered triplet $\bar{t}=\pm\la t_{n,x}\ra$ 
which determines the AF order parameter (polarised along x) and homogeneous 
triplet $\bar{t_0}=\la t_{n,z}\ra$ which determines the magnetisation
caused by the external field along z-direction. For technical reasons it is 
of advantage to transform to circular polarised transverse triplet coordinates (u$_n$,d$_n$) with
respect to the field direction (z axis). The transformation and its inverse are given by
\bea
u_n&=&-\frac{1}{\sqrt{2}}(t_{nx}-it_{ny}) ;\qquad t_{nx}=-\frac{1}{\sqrt{2}}(u_{n}-d_{n}) ,\no\\
d_n&=&\;\frac{1}{\sqrt{2}}(t_{nx}+it_{ny}) ;\qquad t_{ny}=-\frac{i}{\sqrt{2}}(u_{n}+d_{n}) .
\label{CIRC}
\eea
In circular triplet coordinates the Hamiltonian may then be written as
\bea
\HH_0&=&\sum_n\bigl\{
[-\frac{J_x}{4}(2+\Delta)+\mu_n]s_n^\dagger s_n+
[\frac{J_x}{4}(2-\Delta)+\mu_n]t_{nz}^\dagger t_{nz}+\no\\
&&[\frac{J_x}{4}\Delta+\gamma_++\mu_n]u_n^\dagger u_n+
[\frac{J_x}{4}\Delta-\gamma_++\mu_n]d_n^\dagger d_n+
\gamma_-(s_n^\dagger t_{nz}+t_{nz}^\dagger s_n) -\mu_n \bigr\} ,
\label{HAM0}
\eea
and 
\bea
\HH^{(2)}_1&=&-\frac{t}{4}\sum_{\langle n,m\rangle}[(s_n^\dagger s_m^\dagger(u_nd_m+d_nu_m)+h.c.)
-(s_n^\dagger s_m(u_nu_m^\dagger+d_nd_m^\dagger)+h.c.)]\no\\
&&+\frac{\delta t}{4}\sum_{\langle n,m\rangle}[(s_n^\dagger s_m^\dagger t_{nz}t_{mz}+h.c.)
+(s_n^\dagger s_m t_{nz}t^\dagger_{mz}+h.c.)] .
\label{HAM1CI}
\eea
Separating the mean values and the corresponding fluctuations
the Fourier components of singlet and triplet operators are then given
by 
\bea 
s_k&=&\sqrt{N}\bar{s} ,\no\\ 
u_{k}&=&\sqrt{N}\bar{u}\delta_{k,Q} + \hat{u}_{k} ,\no\\ 
d_{k}&=&\sqrt{N}\bar{d}\delta_{k,Q} + \hat{d}_{k} ,\\ 
t_{k,z}&=&\sqrt{N}\bar{t}_0\delta_{k,0}+\hat{t}_{k,z} .\no
\label{SEPARATE}
\eea 
Here we assumed that if AF order appears it will be of the in-plane 
type ($\perp \hat{\mathbf{z}},\mathbf{H}$). Therefore the case of AF order
will only be considered for $\delta\leq 1$.
Instead of $\bth$ it will later be convenient to use $\btau$ 
which is defined by the relation $\bth=\bs\btau$. 
In the paramagnetic phase at zero field only $\bs$ will be different from zero with a value close to
one \cite{Langari06}. As the field increases $\bs$ will decrease and simultaneously the triplet amplitude
$\bth$ associated with the uniform induced moment increases. If an AF transition occurs, $\bs$ will further
decrease and u,d will increase accordingly with the field. The constraint $\bs^2+\bth^2+(|\bu|^2+|\bd|^2)\simeq 1$ is always respected for any field strength. In the limit h/J$_x$ $\gg$1 the amplitudes $\bs$ and $\bth$ become asymptotically equal. On the mean field level this signifies that the ground state of the local level scheme becomes an equal amplitude mixture of the two states with S$_z^t$=0 which show the level repulsion in Fig.~\ref{Fig1}. Therefore it is possible to use the zero-field singlet-triplet bosons in Eq.~(\ref{BOND}) as a basis even in the present finite field problem. On the mean field level, the physical constraint automatically takes care of the change in the local ground state wave function. Inserting the above expressions into the Hamiltonian of Eqs.~(\ref{HAM0},\ref{HAM1CI}) one obtains a
bilinear form in the triplet fluctuation operators ($\hat{u}_k$, $\hat{d}_k$, $\hat{t}_{kz}$)    
which may be diagonalised with two separate Bogoliubov transformations given for the z-polarisation by
\bea
a_{k}=\cosh(\theta_{k,z})\hat{t}_{k,z} 
+\sinh(\theta_{k,z})\hat{t}_{-k,z}^{\dagger}, \nonumber \\
a_{-k}^{\dagger}=\sinh(\theta_{k,z})\hat{t}_{k,z}
+\cosh(\theta_{k,z})\hat{t}_{-k,z}^{\dagger},
\label{BGLZ}
\eea
and for the two circular polarised triplets by
\bea
\label{BGLPER}
A_{k}&=&\cosh(\theta_{k,\perp})\hat{u}_{k} 
+\sinh(\theta_{k,\perp})\hat{d}_{-k}^{\dagger},\no\\
B_{-k}^{\dagger}&=&\sinh(\theta_{k,\perp})\hat{u}_{k}
+\cosh(\theta_{k,\perp})\hat{d}_{-k}^{\dagger},\no\\
B_{k}&=&\cosh(\theta_{k,\perp})\hat{d}_{k} 
+\sinh(\theta_{k,\perp})\hat{u}_{-k}^{\dagger},\\
A_{-k}^{\dagger}&=&\sinh(\theta_{k,\perp})\hat{d}_{k}
+\cosh(\theta_{k,\perp})\hat{u}_{-k}^{\dagger}.\no
\eea
The transformation angles $\theta_{k,z},\theta_{k,\perp}$ are obtained from the 
diagonalisation conditions  
\be
\tanh 2\theta_{k,z}=\frac{2 f_z(k)}{d_z(k)};\qquad
\tanh 2\theta_{k,\perp}=-\frac{2f_\perp(k)}{d_\perp(k)}
\label{ANGLES}
\ee
where the longitudinal auxiliaryy functions f$_z$ and d$_z$ are defined by
\bea
f_z(k)&=&\frac{t \bar{s}^2}{4} \delta \gamma(k); 
\hspace{5mm}d_z(k)=\mu+\frac{2J_x-J_z}{4}+
\frac{t \bar{s}^2}{2} \delta \gamma(k).
\label{FDZ}
\eea
and for the in-plane case  the auxiliary functions f$_\perp$ and d$_\perp$ are given by
\bea
 f_\perp(k)&=&\frac{t \bar{s}^2}{4} \gamma(k); 
\hspace{5mm}
d_{u,d}(k)=\mu+\frac{J_z}{4}+
\frac{t \bar{s}^2}{2} \gamma(k)\pm\gamma_+, \nonumber \\
d_\perp(k)&=&\frac{1}{2}(d_u+d_d)=\mu+\frac{J_z}{4}+
\frac{t \bar{s}^2}{2} \gamma(k).
\label{FDP}
\eea
Here $\gamma(k)=\sum_{i=1}^D\cos(k_i)$  with $\gamma(0)=\frac{z}{2}$ denotes the n.n. structure
factor in dimension $D=\frac{z}{2}$ ($z$ = coordination number of the
simple cubic lattice). It should not be confused with the Zeeman energies $\gamma_\pm$ defined in Sect.~\ref{sec2a}. The Bogoliubov transformations in Eq.~(\ref{BGLZ}) and Eq.~(\ref{BGLPER}) yield 
the diagonalised bilinear Hamiltonian 
\be
\HH_{mf}=E_0+\sum_k 
[\Omega_A(k)A^\dagger(k)A(k) + \Omega_B(k)B^\dagger(k)B(k)+\Omega_z(k) a_{k}^{\dagger}a_{k}] .
\label{HMF}
\ee
where the  triplet mode frequencies $\Omega_\alpha(k)$ ($\alpha$=A,B,z) in the mean field 
Hamiltonian are given by 
\bea
\label{MODEFREQ}
\Omega_{A}(k)&=&\omega_{A}(k)+\gamma_+ ,\no\\
\Omega_{B}(k)&=&\omega_{B}(k)-\gamma_+ ,\no\\
\Omega_{z}(k)&=&\omega_{z}(k) ,\\
\omega_{A}(k)=\omega_B(k)&=&[d_{\perp}(k)^2-4f_{\perp}(k)^2]^\frac{1}{2} 
\equiv\omega_\perp(k),\no\\
\omega_{z}(k)&=&[d_{z}(k)^2-4f_{z}(k)^2]^\frac{1}{2} .\no
\eea
The excitation energies $\Omega_\alpha(k)$ depend on the field both explicitly through $\gamma_+$ 
(for $\alpha$=A,B) and implicitly via $\omega_\alpha$ (k) which is determined by the field dependent 
singlet and triplet amplitudes. In the nonmagnetic phase ($\bu,\bd,\bth$=0) the singlet-triplet excitation gap is 
given by the minimum excitation energy at the incipient ordering wave vector. For D=2 this is at Q=($\pi,\pi$), explicitly $E_g=min\{\Omega_\alpha(Q),\alpha =A,B,z\} $. In the approach from the nonmagnetic side the quantum critical line t$_c$(h) is then defined by the vanishing of E$_g$.

The ground state energy E$_0$ is a function of three control parameters 
(t/J$_x$,$\Delta$,$\delta$), four singlet-triplet expectation values 
($\bar{s},\bar{t}_0,\bar{u},\bar{d}$) and the chemical potential $\mu$. Writing the
transverse mean values explicitly in terms of amplitudes  and phases according to
\be 
\bar{u}=u\exp(i\Phi_u), \hspace{5mm} \bar{d}=d\exp(i\Phi_d) ,
\label{UDPHASE}
\ee
the ground state energy can be written as
\bea
\label{GSEN}
E_0(\frac{t}{J_x},\Delta,\delta;\bs,\bu,\bd,\bth)
&=&N[-\frac{1}{4}(2J_x+J_z)\bs^2+\mu\bs^2-\mu+\no\\
&&(\frac{J_z}{4}+\mu-\frac{1}{4}zt\bs^2)(u^2+d^2)+\gamma_+(u^2-d^2)
+\frac{z}{2}t\bs^2ud\cos(\Phi_u+\Phi_d)\\
&&+(\frac{1}{4}(2J_x-J_z)+\mu+\frac{1}{2}z\delta t\bs^2)\bth^2
+2\gamma_-\bs\bth]\no\\
&&+\frac{1}{2}\sum_{k\alpha}[\omega_k^\alpha-d_\alpha] .\no
\eea
As a first step we determine the triplet condensate phases ($\Phi_u,\Phi_d$) by
minimization. Since only one term depends on the phases we obtain the extremal condition
$\sin(\Phi_u+\Phi_d$)=0. The minimum must also satisfy $\cos(\Phi_u+\Phi_d)<0$. This is achieved 
for $\Phi_u+\Phi_d=n\pi$ with n an odd integer, i.e. the sum is only determined modulo 2$\pi$. 
Since there is no condition for the difference of phases one of them is arbitrary. We then 
choose  ($\Phi_u,\Phi_d$)=(0,$\pi$). The remaining continuous degeneracy with respect to the phase difference is a signature of the Goldstone mode which is present throughout the AF phase.
%
\begin{figure}[tbc]
\begin{center}
\includegraphics[width=8.0cm,angle=0]{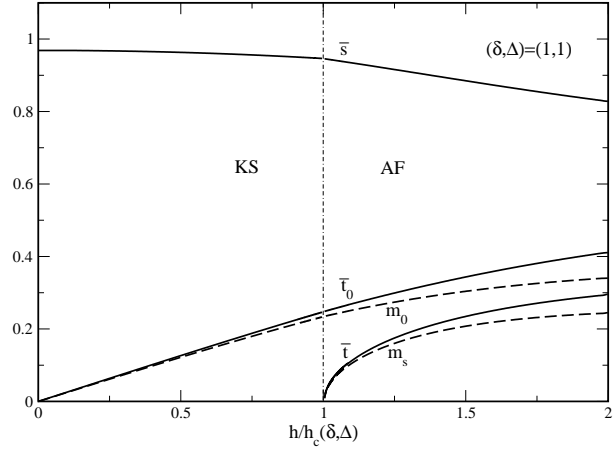}
\caption{Singlet ($\bs$) and triplet ($\bth,\bt$)  amplitudes and their associated uniform ($m_0=\bs\bth$)  and staggered ($m_s=\bs\bt$) moments as function of field strength normalised to the critical field in the KS and AF region. Here h$_c$($\delta,\Delta$)/J$_x$=0.55. At the critical field $(\bth/\bs) = 0.25$. This case with t/t$_c$($\delta,\Delta$)=0.6 (t/J$_x$ =0.517) corresponds to the the upper curve of Fig.~\ref{Fig4} (top). Here ($g_s,g_\tau$)=(2.0).}
\label{Fig2}
\end{center}
\end{figure}
%
\section{Selfconsistent equations for singlet and triplet amplitudes
and magnetisation}
\label{sec4}
The minimisation of the ground state energy E$_0$ in Eq.~(\ref{GSEN})
leads to selfconsistent coupled equations for the condensate
amplitudes $\bs,\bu,\bd,\bth$ and the chemical potential $\mu$. Their
structure is slightly different in the nonmagnetic ($\bt$=0) and
magnetic ($\bt > 0$) case, therefore we write them explicitly for both.
 For a convenient expression of the selfconsistency
equations we define the quantity $\btau$ by the
relation $\bth\equiv\btau\bs$. Furthermore we introduce the Brillouin zone integrals
F$_\alpha$ and G$_\alpha$ ($\alpha$=A,B,z) given by
\bea
F_\alpha&=&\frac{1}{N}\sum_k\frac{d_\alpha}{\omega_{k\alpha}} ,\no\\
G_\alpha&=&\frac{t}{N}\sum_k\frac{d_\alpha(k)-2f_\alpha(k)}{\omega_{k\alpha}}\gamma_k^\alpha .
\label{FG}
\eea
which appear in the extremal conditions when differentiating the last
term in E$_0$ with respect to $\mu$ or $\bs$ respectively. Here we have
defined $\gamma_k^\perp\equiv\gamma_k^{A,B}=\gamma_k$ and
$\gamma_k^z=\delta\gamma_k$. Note that even in the case of finite
field ($\gamma_\pm \neq 0$) it is the frequency $\omega_{k\alpha}$ and not the mode
frequency $\Omega_{k\alpha}$ which appears in the expressions for F$_\alpha$ and
G$_\alpha$. \\
%
\begin{figure}
\begin{center}
\includegraphics[width=8cm,angle=0]{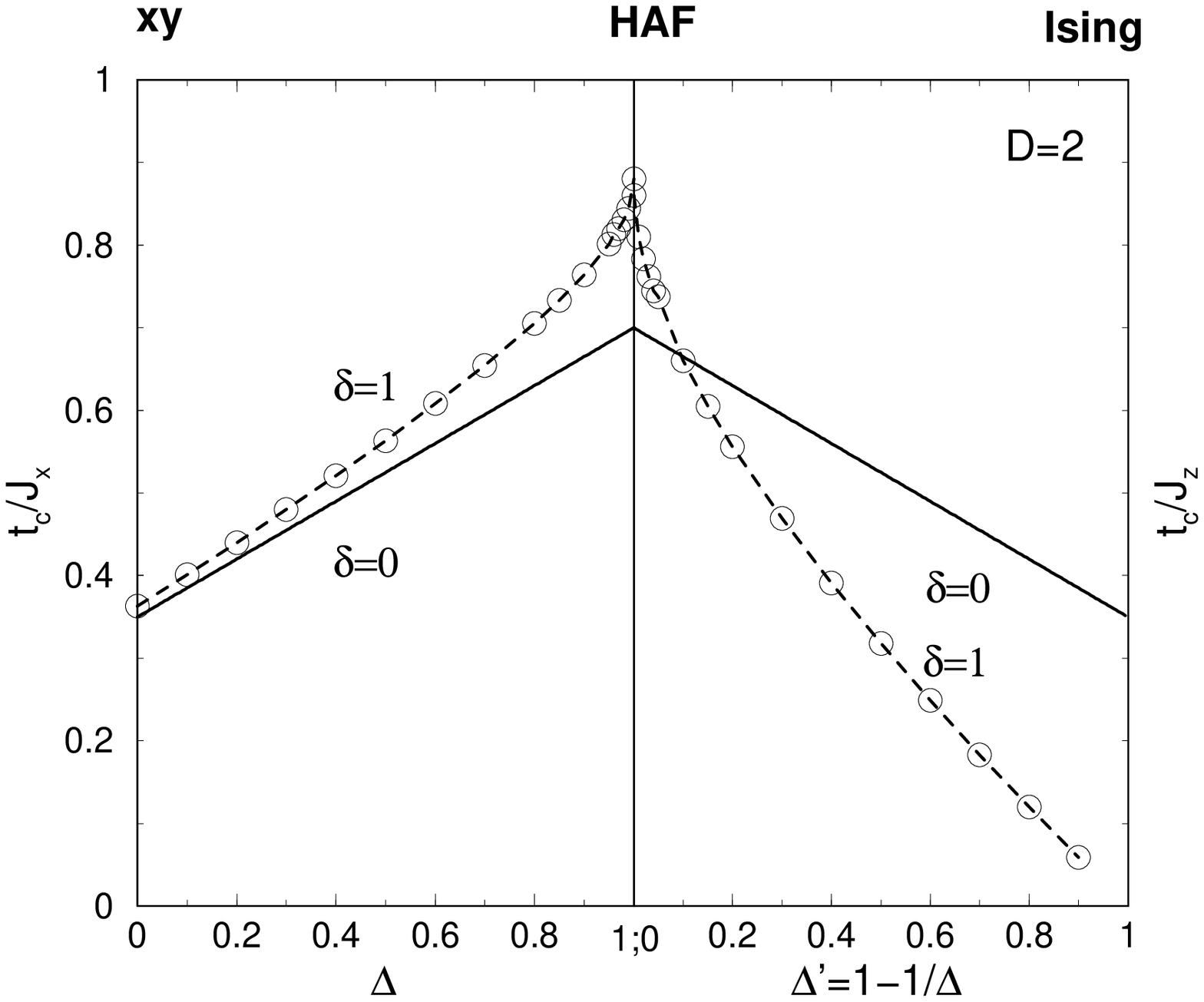}\\
\vspace{1cm}
\includegraphics[width=8cm,angle=0]{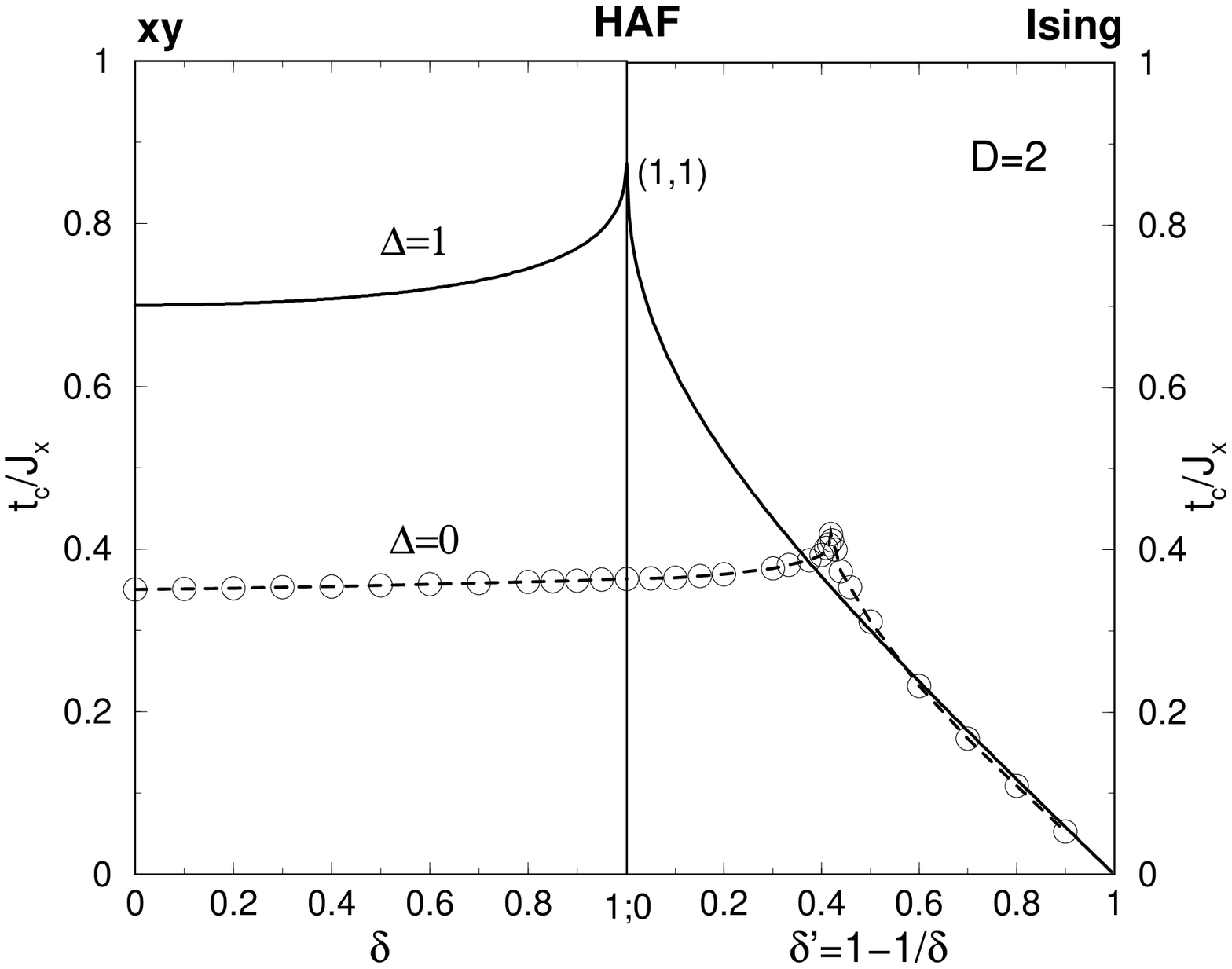}
\caption{ Critical hopping strength t$_c$($\delta$,$\Delta$) for quantum phase transition between Kondo singlet (t$<$t$_c$) 
and antiferromagnetic  (t$>$t$_c$) phases at zero field. For ($\delta$,$\Delta$)=(0,$\Delta$) or ($\delta$,1)  t$_c$ is obtained from a closed analytical expression (Eq.(25) in Ref.~\onlinecite{Langari06}) (full lines), for the other cases it is calculated from the zero-field version of Eq.~(\ref{PARA}) (dashed lines). Top: t$_c$-dependence on anisotropy $\Delta$ of local spins . In the Ising case ($\Delta'\rightarrow$ 1) t$_c$  for $\delta$=0 does not vanish because the AF order is due to the mixing with the doublet separated by a finite excitation gap. Bottom:  t$_c$-dependence on anisotropy $\delta$ of interacting spins. The value t$_c$(0,1)/J$_x$=0.7 agrees with results from MC simulatons for the genuine Kondo lattice model in 2D. 
\cite{Wang94,Assaad99}}
\label{Fig3}
\end{center}
\end{figure}
%

{\it Nonmagnetic Kondo spin singlet phase in external field}\\

In this case the transverse triplet amplitudes vanish, i.e., u=d=0 and the minimization 
of E$_0$ with respect to ($\bs,\bth\equiv\bs\btau,\mu$) leads to the selfconsistent set of equations 
\bea
\label{PARA}
\bs^2&=&\frac{\frac{1}{2}(5-\sum_\alpha F_\alpha)}{1+\btau^2} ,\no\\
\btau&=&-\frac{2\gamma_-}{2\mu+\frac{1}{2}(2J_x-J_z)+z\delta t\bs^2} ,\\
\mu&=&\frac{1}{2}[(J_x+\frac{1}{2}J_z)-\frac{1}{2}\sum_\alpha G_\alpha
-2\gamma_-\btau -z\delta t\bs^2\btau^2] .\no
\eea
Here $\bs, \mu$ are found by iteration which determines $\btau$ and
hence $\bth$ completely via the second equation above. For zero field, i.e. $\gamma_-=0$,
the induced $\btau$ vanishes and then Eq.~(\ref{PARA}) reduces to the paramagnetic case
of Ref.[\onlinecite{Langari06}].\\  
%
%
\begin{figure}
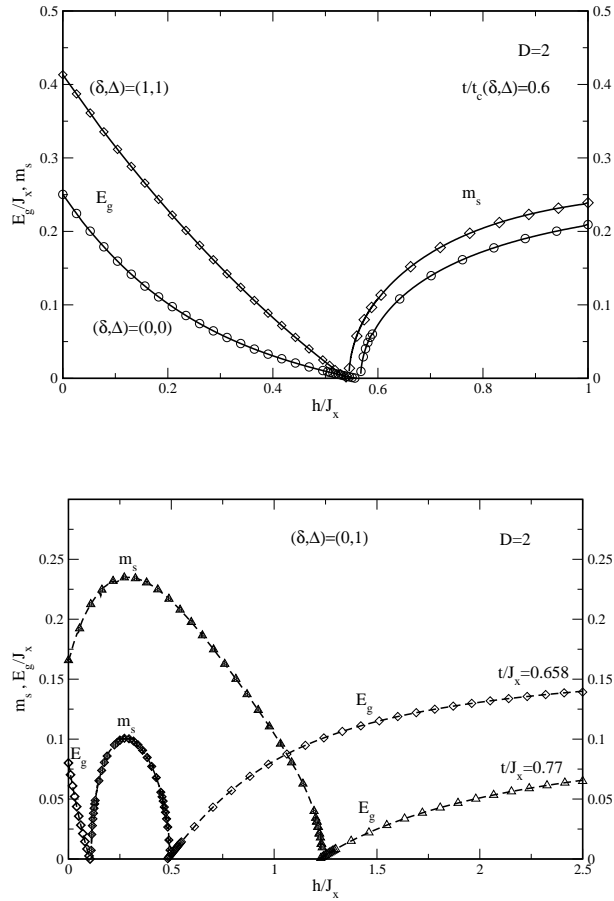

\begin{center}
\includegraphics[width=8.0cm,angle=0]{Fig4a}\\
\vspace{1cm}
\includegraphics[width=8.0cm,angle=0]{Fig4b}
\caption{In these figures three possible cases for quantum phase transitons between Kondo singlet KS(E$_g$ = singlet gap) and
 antiferromagnetic AF(m$_s$ = staggered moment) phases as function of increasing field strength h are shown: (i) field induced AF with sequence  KS -AF. (ii) destruction of AF and field induced gap opening corresponding to the sequence KS-AF. (iii) reentrant behaviour corresponding to KS-AF-KS.
Top:  Singlet-triplet excitation gap E$_g$ and AF order parameter m$_s$ as function of external field for the two cases 
$(\delta,\Delta)$ =(0,0) (circles) and $(\delta,\Delta)$ =(1,1) (diamonds) as function of the external field. A subcritical scaled value
 t/ t$_c$($\delta,\Delta$) = 0.6 is used in both cases (t$_c$(0,0)/J$_x$=0.350, t$_c$(1,1)/J$_x$=0.862). Here (g$_s$,g$_\tau$)=(2,0) is used. This plot corresponds to the field induced AF case (i).
Bottom: Similar plot for  $(\delta,\Delta)$ = (0,1) and an above critical  t/ t$_c$($\delta,\Delta$) = 1.1  (t/J$_x$=0.77) and subcritical  t/ t$_c$($\delta,\Delta$) = 0.94  (t/J$_x$=0.658).  In the former case (ii)  one has the sequence AF-KS of phases  ($\triangle$) and  in the latter (iii) a reentrant situation with KS-AF-KS sequence is observed ($\diamond$).}
\label{Fig4}
\end{center}
\end{figure}
%

{\it Antiferromagnetic phase in external field}\\

The additional minimization with respect to u and d amplitudes leads to a direct
relation for the chemical potential $\mu$ in terms of the singlet amplitude
$\bs$ which is a generalised version of the zero-field AF case \cite{Langari06}:
\be
\mu(h)=\frac{1}{4}zt\bs^2-\frac{J_z}{4}+[(\frac{1}{4}zt\bs^2)^2+\gamma_+^2]^\frac{1}{2} .
\ee
At this stage it is convenient to introduce an auxiliary parameter $\kappa$ which controls the 
effect of the external field on the boson amplitudes; it is defined by
\be
\kappa(h)=[1+(\frac{\gamma_+}{\frac{z}{4}t\bs^2})^2]^\frac{1}{2} +\frac{\gamma_+}{\frac{z}{4}t\bs^2}.
\label{KAPPA}
\ee
Obviously $\kappa$(0)=1 and $\kappa(h)\leq 1$ since $\gamma_+<0$. Using the expression for the zero-field chemical potential 
$\mu_0$=$\mu(0)$ we may also write 
\bea
\mu(h)&=&\mu_0+\frac{z}{4}t\bs^2\frac{(\kappa-1)^2}{2\kappa},\no\\
\mu_0&=&\frac{z}{2}t\bs^2-\frac{1}{4}J_z .
\eea
We define a total transverse triplet amplitude $\bt$ by
$\bt^2=\frac{1}{2}(u+d)^2$ which should not be confused with the hopping
matrix element t of the Hamiltonian.  The minimization with respect to
u,d leads to a relation between the two amplitudes from which we
obtain
\be
u^2+d^2=2\bt^2\frac{1+\kappa^2}{(1+\kappa)^2}; \hspace{5mm} 
u^2-d^2=2\bt^2\Bigl(\frac{1-\kappa}{1+\kappa}\Bigr),
\label{UD1}
\ee
or equivalently
\be
u^2=\frac{2\bt^2}{(1+\kappa)^2}; \hspace{5mm}
d^2=\frac{2\bt^2\kappa^2}{(1+\kappa)^2}=\kappa^2u^2.
\label{UD2}
\ee
The remaining set of singlet and triplet amplitudes ($\bs,\bt,\bth=\bs\btau$) 
is again determined by the solution of three coupled equations:
\bea
\label{AF}
\bs^2&=&\frac{\frac{1}{2}(5-\sum_\alpha F_\alpha) -(u^2+d^2)}{1+\btau^2} ,\no\\
\btau&=&-\frac{2\gamma_-}{2\mu+\frac{1}{2}(2J_x-J_z)+z\delta t\bs^2}
\equiv -2\gamma_-\hat{\tau}_0 ,\\
\bt^2&=&\frac{1}{zt}[2\mu-\frac{1}{2}(2J_x+J_z)+2\gamma_-\btau +
\frac{1}{2}\sum_\alpha G_\alpha]+\delta\bs^2\btau^2 .\no
\eea
These amplitudes are related to the spin expectation values by
\bea
\la S_z\ra&=&\bs\bth+\frac{1}{2}(u^2-d^2)\no\\
\la \tau_z\ra&=&-\bs\bth+\frac{1}{2}(u^2-d^2)
\label{SPINAV}
\eea
For zero field ($\btau$=0) the system of equations in Eq.~(\ref{AF})  may be shown to reduce to
the one already discussed in Ref.[\onlinecite{Langari06}]. In this case $\kappa$=1 and
u=d, i.e., circular polarised triplet modes with equal amplitudes. 

Finally we discuss the uniform magnetisation M$_0$  and the staggered magnetisation 
M$_s$=M$_B$=-M$_A$ associated with AF order. By definition
\bea
M_0&=&g_s\langle S_z\rangle +g_\tau\langle\tau_z\rangle\no\\
M_s&=&g_s\langle S^B_x\rangle +g_\tau\langle\tau^B_x\rangle.
\label{UNSTA}
\eea
Here the staggered moment is assumed to be polarised along x. 
Using the expressions in Eq.~(\ref{SPINAV})  for $\la S_{z}\ra$, $\la\tau_{z}\ra$ and similar ones for 
$\la S_{x}\ra$, $\la\tau_{x}\ra$  we obtain with the help of Eqs.~(\ref{KAPPA},\ref{UD1})
\bea
M_0&=&\bs\bth[2g_- +2g_+\lambda\bigl(\frac{\bt^2}{\bs\bth}\bigr)]\simeq 2g_-\bs\bth \equiv 2g_-m_0,\no\\
M_s&=&\bs\bt[2g_- -2g_+\lambda\bigl(\frac{\bth}{\bs}\bigr)]\simeq 2g_-\bs\bt\equiv 2g_ -m_s .
\label{MAGNET}
\eea
Here we used  $g_\pm=\frac{1}{2}(g_s\pm g_\tau)$ and defined
\bea
\lambda=\frac{1-\kappa}{1+\kappa} .
\eea
When both staggered and uniform component are nonzero the total magnetic moment is canted. The canting angle $\alpha$, counted from the xy plane is then given by
\bea
\label{CANTING}
\tan\alpha=\Bigl(\frac{\bth}{\bt}\Bigr)
\frac{g_- +g_+\lambda\bigl(\frac{\bt^2}{\bth\bs}\bigr)}
{g_--g_+\lambda\bigl(\frac{\bth}{\bs}\bigr)} .
\eea
 %
\begin{figure}
\begin{center}
\includegraphics[width=8.0cm,angle=0]{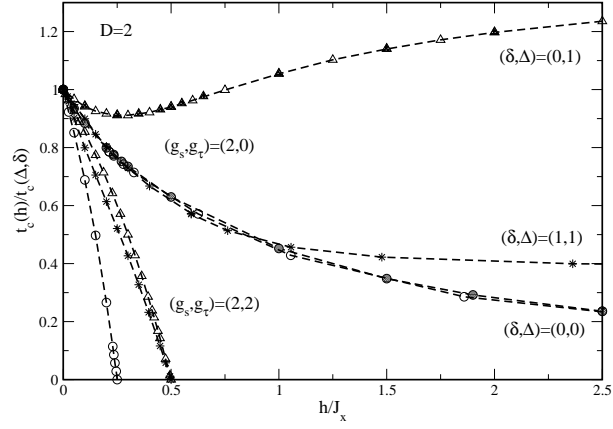}
\caption{t-h Phase diagram (critical t$_c$(h) curves) for various cases of anisotropies ($\delta$,$\Delta$) and g-factors
(g$_s$,g$_\tau$). Here circles ($\circ$) correspond to ($\delta$,$\Delta$)=(0,0)   stars ($\star$) to  ($\delta$,$\Delta$)=(1,1) and 
triangles ($\triangle$) to ($\delta$,$\Delta$)=(0,1). Full symbols are obtained by m$_s$=0 from the AF side, open symbols by E$_g$=0 from the Kondo singlet side. For unequal g-factors  the Heisenberg case shows saturation at $t_c/J_x\simeq$ 0.4 at larger fields, in the Ising  case t$_c$ continues decreasing.  For the mixed case ($\delta$,$\Delta$)=(0,1) (the genuine KN model in 2D) and unequal g-factors the t$_c$(h) is nonmonotonic. In each case the KS phase is below and the AF phase above the corresponding t$_c$(h) curve. A decreasing  t$_c$(h) leads to the phase sequence KS-AF (field- induced AF) whereas an increasing t$_c$(h) entails the opposite sequence AF-KS. In the nonmonotonic region the reentrance behaviour KS-AF-KS is observed. For comparison see Fig.~\ref{Fig4}. For equal g-factors  in all  cases t$_c$ decreases  monotonically and  vanishes at rather small fields.}
\label{Fig5}
\end{center}
\end{figure}
%
The terms $\sim\lambda$ are higher order corrections, e.g. $\sim(h/J_x)^2$ for
M$_s$. For moderate fields they may be neglected. In Sec.~\ref{sec5} (Fig.~\ref{Fig4}) we only show the
main contribution $m_s\equiv \bs\bt$ in accordance with Ref. \onlinecite{Langari06}.  
In Eq.~(\ref{CANTING}) a similar small field approximation may be employed for the canting angle leading to 
 $\alpha\simeq\tan^{-1}(\bth/\bt) $. For  equal g-factors g$_s$=g$_\tau$  (g$_-$=0 and $\bth$=0) one has a special case:  M$_s$ = 0 due to moment compensation of local and itinerant spins , although this is still a  phase with an AF order parameter m$_s$=$\bs\bt$. The uniform moment M$_0$ is now given by $M_0=2g_+\lambda\bt^2$, oriented parallel to the external field ($\alpha=\pi/2$).\\[0.5cm]

 {\it The ground state energy: condensation vs. fluctuation}\\

It is instructive to rewrite the ground state energy in a different way which
renders a clearer interpretation of its individual contributions. The total mean
field Hamiltonian in Eq.~(\ref{HMF}) may be rewritten as 
\be
\HH_{mf}=\tilde{E}_0+\sum_{\alpha=z,A,B}\sum_{k}\Omega_\alpha(k)[n_\alpha+\frac{1}{2}] ,
\label{HMFTIL}
\ee
where n$_\alpha$ is the occupation number of bosons for each mode. The total ground state energy
is then given by
\be
E_0=\tilde{E}_0+\frac{1}{2}\sum_{k\alpha}\Omega_\alpha(k) ,
\label{GSENTOT}
\ee
where $\tilde{E}_0(h)$ consists of three parts according to
\bea
\label{GSENTIL}
\tilde{E}_0/N&=&-[\frac{1}{4}(2J_x+J_z)(\bs^2+\frac{1}{2})+\mu(\frac{5}{2}-\bs^2)]\no\\
&&+\frac{z}{2}t(\kappa-1)^2\bs^2\bt^2+(\frac{1}{4}(2J_x-J_z)+\mu+\frac{z}{2}\delta t\bs^2)\bth^2\\
&&+2\gamma_-\bs\bth .\no
\eea
Here the first part ($\tilde{E}_0$) in Eq.(\ref{GSENTOT}) is the condensate
contribution. The second always positive term is due to triplet quantum fluctuations.
According to Eq.(\ref{GSENTIL}) $\tilde{E}_0$ is composed of three
parts: The first (negative) contribution is due to the singlet
formation. The second (positive) part is due to the field induced
triplet polarisation. The last term is the Zeeman energy
contribution. For zero field only the first term in Eq.(\ref{GSENTIL})
is present \cite{Langari06}. The ground state with $E_0<0$ will be determined by the
competition of these terms. For example when h = 0 the quantum critical
point where AF order appears will be determined by the balance of two
terms: The negative singlet formation energy and the positive energy of triplet
quantum fluctuations.

\section{Influence of higher order terms}
\label{sec4a}

The transformation of the Hamiltonian in Eq.~(\ref{HAM}) to bond operator variables also creates third and fourth order terms in the triplet operators. They have been neglected in the previous analysis based on Eqs.~(\ref{HAML},\ref{HAM1}). 
In fact it was argued in  \cite{Langari06}  that third order terms do not contribute to the ground state energy and fourth order terms are quantitatively negligible. The latter was also found for the related bilayer Heisenberg model \cite{Matsushita99}.
 Hence higher order terms have no influence on the zero-field quantum critical properties. In this section we investigate to what extent this is still justified in the presence of an external field. Indeed in mean field approximation used here the higher order terms lead to field induced effective bilinear triplet terms which have to be added to the genuine bilinear term in Eq.~(\ref{HAM1CI}) which is present already at zero field. The quantitative influence of higher order terms on t$_c$(h) or h$_c$(t) is controlled by the ratio $(\bth/\bs)$ at the critical field h$_c$ where $\bth(h_c)$ is the field induced triplet amplitude. If it is still moderate compared to the singlet amplitude $\bs(h_c)$ then higher order terms have negligible influence. The field dependence of these amplitudes as obtained from Eqs.~(\ref{PARA},\ref{AF}) is shown in Fig.~\ref{Fig2} for a typical case.

The contribution of third order in t$_{n\alpha}$ (n= site, $\alpha$=x,y,z) to the interaction term  is given by \cite{Langari06}:
\bea
\HH^{(3)}_1&=&\frac{i t}{4}\sum_{\la n,m\ra}[
(s_n^{\dagger}t_{nx}+t_{nx}^{\dagger}s_{n})(t^{\dagger}_{my} t_{mz}-t^{\dagger}_{mz} t_{my})
+(s_n^{\dagger}t_{ny}+t_{ny}^{\dagger}s_{n})(t^{\dagger}_{mz} t_{mx}-t^{\dagger}_{mx} t_{mz})
\nonumber \\
&+&\delta(s_n^{\dagger}t_{n,z}+t_{n,z}^{\dagger}s_{n})(t^{\dagger}_{m,x} t_{m,y}-t^{\dagger}_{m,y} t_{m,x})].
\label{HAM3}
\eea
which has to be added to Eq.~(\ref{HAMMU}). We are only interested in the influence of this term on the quantum critical lines h$_c$(t) or t$_c$(h) to be discussed in Sect.~\ref{sec5}. As explained there this may be achieved both from the paramagnetic (KS) and AF side of the critical line. Here we consider only the former for reasons explained at the end of the section. In this case the circular triplet amplitudes $\bar{u}=\bar{d}\equiv 0$. The mean field approximation to H$^{(3)}_1$ then contains only terms proportional to $\bar{t_0}$ and no constant contribution to E$_0$ appears. Transforming these terms to circular triplet coordinates  one finally obtains another  bilinear contribution with
\bea
\HH^{(3)}_{1mf}
=-\frac{t}{4}z\delta\bs\bth\sum_n(u^\dagger_nu_n-d^\dagger_nd_n)
+\frac{t}{2}\bs\bth\sum_{\la nm\ra}(u^\dagger_nu_m-d^\dagger_nd_m).
\label{HAM3MF}
\eea
This contribution is field induced since for small fields $\bth(h)\sim h$. 
The first single site term in Eq.~(\ref{HAM3MF}) has the same structure as the $\gamma_+$ part of the Zeeman term and may be accommodated by a simple (nonlinear) rescaling of the applied field  such that $\gamma_+$ is replaced by $\tilde{\gamma}_+$ according to
\bea
\label{HCSCALE}
\tilde{\gamma}_+(h)=-g_+hf_s(h); \qquad f_s(h)=1+\frac{z}{4}\delta t\frac{\bs(h)\bth(h)}{h}
\eea
Here $f_s(h)$ is the rescaling function for the applied field. Note that for $\delta =0$ one has $f_s(h)\equiv 1$, i.e. no rescaling will occur in this case. The second contribution in Eq.~(\ref{HAM3MF}) is an 
interaction to which the same (transverse) Bogoliubov transformation as before may be applied. These terms then have a simple
effect: In Eq.~(\ref{FDP})  one has to replace $d_{u,d}(k)\rightarrow d_{u,d}(k)\pm(t/2)\bs\bth\gamma(k)$. However only the 
average $d_\perp(k)=(d_u(k)+d_d(k))/2$ of the auxiliary functions, which is unchanged, enters the expression for $\omega_{A,B}(k)$ in Eq.~(\ref{MODEFREQ}). Therefore $\omega_{A,B}(k)$ is also unchanged by the third order contribution, As a result the total energy E$_0$ in Eq.~(\ref{GSEN}) will be exactly the same as before. Also the selfconsistent equations Eq.~(\ref{PARA}) will be unchanged. Note that in these equations  $\gamma_-$ is not rescaled. The only effect of the third order terms is therefore the above rescaling of the external field in the $\gamma_+$ Zeeman term leading to the modified transverse mode frequencies
\bea
\tilde{\Omega}_{A,B}(k)=\omega_{A,B}(k)\pm\tilde{\gamma}_+. 
\eea
Therefore the critical field h$_c$ of the quantum phase transition which is defined as the field where one of the above modes ($\tilde{\Omega}_+(k)$ for $h>0$) vanishes at the AF wave vector {\bf Q}=$(\pi,\pi)$ will be changed by the scaling factor $f_s(h_c)$.  If h$_c^0$  is the critical field without third order terms then approximately $h_c=h^0_c/f_s(h_c^0)$ is the critical field with the effect of third order terms included.
This is only a quantitative modification which we will discuss in Sec.~\ref{sec5} in connection with Fig.~\ref{Fig5}. The qualitative topology of the phase diagram will not be changed by the third order term. We note again that in the special case $(\Delta,\delta)$=$(\Delta,0)$ the third order term does not have any effect at all because $f_s\equiv 1$. Furthermore for all cases g$_s$=g$_\tau$ there is no field induced triplet amplitude, i.e. $\bth\equiv 0$ according to Eq.~(\ref{PARA}) leading again to  $f_s\equiv 1$. We conclude that the influence of third order terms is not important for the field induced quantum critical behaviour .\\

Now we discuss the effect of terms which are of fourth order in the triplet operators. They are given by \cite{Langari06} 
\bea
\HH^{(4)}_1&=&\frac{-t}{4}\sum_{\la n,m\ra}\Big((t^{\dagger}_{ny} t_{nz}-t^\dagger_{nz}t_{ny})
(t^{\dagger}_{my} t_{mz}- t^\dagger_{mz}t_{my})
+(t^{\dagger}_{nx} t_{nz}- t^\dagger_{nz}t_{nx})(t^{\dagger}_{mx} t_{mz}-t^\dagger_{mz}t_{mx}) \no\\
&+&\delta(t^{\dagger}_{nx} t_{ny}- t^\dagger_{ny}t_{nx})
(t^{\dagger}_{mx} t_{my}- t^\dagger_{my}t_{mx})\Big).
\eea
The mean field approximation to $H^{(4)}$ for the nonmagnetic case ($\bu=\bd=0$ ) does not produce a constant term but leads to an effective bilinear contribution in circular triplet coordinates
\bea
\HH^{(4)}_{1mf}=\frac{t}{2}\bth^2\sum_{\la nm\ra}
[u_n^\dagger u_m + d_n^\dagger d_m
+\frac{1}{2}(u_nd_n+d_nu_m+u_n^\dagger d_m^\dagger +d_n^\dagger u_m^\dagger)].
\eea
Again this is a field-induced bilinear contribution with $\bth(h)\sim h^2$ for small fields. Due to $\bu, \bd=0$ it has no longitudinal part. It may be diagonalised by the same transverse Bogoliubov transformation in Eq.~(\ref{BGLPER}) as before. The new mode frequencies including fourth order contributions are simply obtained by the replacement  $\tilde{d}_\perp= d_\perp+(t/2)\bth^2\gamma(k)$ and  $\tilde{f}_\perp = f_\perp-(t/4)\bth^2\gamma(k)$ in Eq.~(\ref{FDP}). This will lead to the modified transverse frequency 
\bea
\tilde{\omega}_{A,B}(k)^2=\omega_{A,B}(k)^2
[1+\bth^2\frac{t\gamma(k)}{d_\perp-2f_\perp}],
\eea
and new mode frequencies $\tilde{\Omega}_{A,B}(k)=\tilde{\omega}_{A,B}(k)\pm\gamma_+$
At the quantum critical point where $\tilde{\Omega}_A(Q)=0$ this may be approximately written as
\be
\tilde{\omega}_{A,B}(k)\simeq\omega_{A,B}(k)
[1+\frac{1}{z}\bigl(\frac{\bth}{\bs}\bigr)^2_c\gamma(k)],
\label{FOURTHC}
\ee
where the index c denotes the values of singlet and triplet amplitudes at the QCP. From Fig.~\ref{Fig2} we can estimate that the prefactor in Eq.(\ref{FOURTHC}) is of the order $\frac{1}{z}\bigl(\frac{\bth}{\bs}\bigr)^2\simeq 10^{-2}$ since at the QCP the triplet amplitude is still quite small and the singlet amplitude basically unchanged from the zero-field value. Therefore the fourth order terms lead to corrections of the order one per cent in the frequencies $\omega_{A,B}(k)$ which determine the last term in the ground state energy of Eq.~(\ref{GSEN}). Indeed the correction to $E_0$ is even smaller since the momentum summation over $\omega_{A,B}(k)$ in the last term of  Eq.~(\ref{GSEN}) leads to a large amount of cancellation because $\gamma(k)$ is positive in one half of the Brillouin zone and negative in the other half. Therefore the selfconsistent equations for $\bs$ and $\bth$ are nearly unchanged and we can conclude that the field induced quantum critical behaviour is  not influenced by the inclusion of fourth order terms. Our analysis shows that the quantum critical lines h$_c$(t)  are only marginally influenced by the presence of higher order terms in one example and have no effect in other cases. However for fields with $h\gg h_c$ one would expect that the higher order terms should lead to non-negligible corrections, at least in the case of third order contributions. From Fig.~\ref {Fig2} one can see that for $h/h_c > 2$ the ratio $(\bth/\bs)$  approaches 0.5 and the bilinear approximation might become inadequate.  In fact, by its very construction the bond-operator method  is a strong coupling theory which assumes the dominance of the singlet state. It cannot be expected to be quantitatively correct for high fields when saturation of moments is approached, i.e. when singlet and triplet amplitudes become equal. In this limit it is more appropriate to start from the polarised canted state and perform a conventional spin wave expansion for the two types of spins.\\

As a consequence of the above analysis we will only consider the genuine bilinear Hamiltonian Eq.~(\ref{HAM1CI}) as used in the previous sections for the following numerical calculations. This restriction has an additional reason: Although the critical $h_c(t)$ or t$_c$(h) may be calculated by searching for the vanishing of the excitation gap E$_g$ from the nonmagnetic singlet side, it is more satisfactory to search also from the AF magnetic side for the vanishing of the staggered magnetisation m$_s$ and confirm the agreement. This cannot be done easily when third and fourth order terms are included. On the AF side the latter lead to a mixing of the longitudinal and transverse modes and a closed analytic form of the Bogoliubov transformation cannot be found. Therefore it is better to include only the bilinear terms and be aware of the trivial effect of field rescaling by the third order terms and the tiny effect of fourth order terms. This standpoint will be adopted for the following 
 numerical analysis.


\section{Numerical solution for excitation gap and AF order parameter, critical field curves and the h-t phase diagram}
\label{sec5}
We shall now discuss the numerical solutions of the selfconsistent Eqs.~(\ref{PARA},\ref{AF}) which describes the AF-KS quantum phase transitions, i.e., the field dependence of the Kondo singlet gap E$_g$  and the field dependence of the staggered magnetisation m$_s$. 
The former was defined below Eq.~(\ref{MODEFREQ}) and the latter in Eq.~(\ref{MAGNET}). We shall only discuss the results for D=2 and $\delta\leq 1$, i.e., the easy xy-plane situation with $\mathbf{m}_s\perp$ c. In this case $\Omega_{A,B}(k)\leq\Omega_z(k)$. 
As shown before the qualitative behaviour of D=3 is similar to D=2 \cite{Langari06}.
The  main purpose is to study the dependence of the quantum critical point t$_c$(h) on the external field, or equivalently the dependence of the quantum critical field h$_c$(t) on the hopping strength t. To check consistency the QCP has been obtained both by variation of t and of h  and both from the paramagnetic (E$_g$=0) as well as the antiferromagnetic (m$_s$=0) side of the QCP.  As mentioned in the previous section this requires restriction to the genuine bilinear Hamiltonian terms.\\
%
\begin{figure}[tbc]
\begin{center}
\includegraphics[width=8.0cm,angle=0]{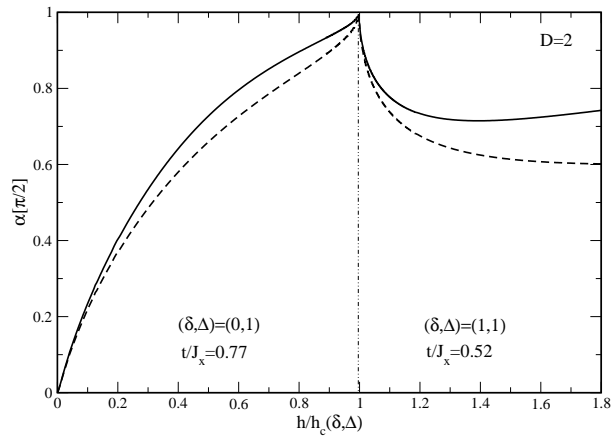}
\caption{Canting angle $\alpha$ of total moment moment (counted from the xy-plane) in the AF phase of two different cases (cf. Fig.~\ref{Fig4}) with (g$_s$,g$_\tau$)=(2,0). Full curve is obtained from the exact expression Eq.~(\ref{CANTING}) and dashed curve from the  approximate expression $\alpha\simeq\tan^{-1}(\bth/\bt)$. At the quantum critical field h$_c$ the staggered moment vanishes and the total moment is aligned with the external field ($\alpha=\pi/2$). Here h$_c$(0,1)/J$_x$=1.23  and  h$_c$(1,1)/J$_x$=0.55.}
\label{Fig6}
\end{center}
\end{figure}
%

Before discussing the field dependent results we show the behaviour of the quantum critical $t_c(\Delta,\delta)$  as function of the anisotropies in the zero field case as a starting point of our analysis. The calculation of  $t_c(\Delta,\delta)$  has been described in Ref.~[\onlinecite{Langari06}]. Here we present the results  for the full range of anisotropies (Fig.~\ref{Fig3}) from the U(1) -xy case  via Heisenberg- SU(2) to Ising-Z$_2$ symmetry. Note the two different scales on the left (J$_x$) and right (J$_z$) half of each part of Fig.~\ref{Fig3}. 
The quantum critical t$_c$ for the KS-AF transition  reaches a singular maximum for the Heisenberg case and generally vanishes for the Ising case, except for $\delta$=0. This may be understood by comparing with Fig.~\ref{Fig1}. For $\Delta\rightarrow\infty$ (i.e. $\Delta'\rightarrow 1$) the two singlets with S$^t_z$=0 become degenerate and an arbitrary small interaction t leads to the AF state. However for $\delta$=0 the S$^t_z$=0 states are not connected by a matrix element of $\HH^{(2)}_1$ and hence AF can occur only via matrix elements between  the singlet ground state and the excited doublet which requires a finite t$_c$/J$_x$ = 0.35 for AF order to occur. \\

The effect of the external field for  (g$_s$,g$_\tau$) = (2,0) is shown in Fig.~\ref{Fig4} (top) for the two cases ($\Delta,\delta$) = (0,0) ($\circ$) and (1,1) ($\Diamond$). Their corresponding  t$_c$ values may be taken from Fig.~\ref{Fig3}. In Fig.~\ref{Fig4} (top)  we choose a subcritical value t/t$_c(\Delta,\delta)$ = 0.6 meaning a nonmagnetic singlet phase with finite E$_g$ exists for zero field. When the field is increased the gap is gradually reduced until it is closed at t$_c$(h$_c$). For the (normalised) t value of 0.6 obviously the difference in critical fields h$_c$ is not significant. However, the qualitative behaviour of the spin gap E$_g$ in the two cases for h$<$h$_c$ is different with a much smaller slope  of E$_g$ in the xy-case as compared to the Heisenberg case. In both cases for h$>$h$_c$ a field induced easy plane AF order parameter appears, i.e., a phase sequence KS-AF for increasing field. The opposite behaviour is observed in the genuine KN case  ($\Delta,\delta$) = (1,0) 
with  (g$_s$,g$_\tau$) = (2,0) in  Fig.~\ref{Fig4} (bottom). For an above- critical value  t/t$_c(0,1)$ = 1.1 (t/J$_x$=0.77) the AF order parameter is gradually suppressed until it vanishes at h$_c$/J$_x$ =1.23 and the Kondo singlet phase appears, i.e., the opposite phase sequence AF-KS is realised. Finally, for slightly subcritical  t/t$_c(0,1)$ = 0.94 (t/J$_x$=0.658)  an interesting  reentrance sequence KS-AF-KS of phases as function of field is observed. It exists only in a narrow range of subcritical values $0.91< t/t_c\ <1.0$. As discussed in Sect.~\ref{sec4a} this behaviour is robust because third order triplet contributions are exactly zero for $\delta$=0. \\

 We may collect the data of  quantum critical fields h$_c$ from numerous calculations such as presented in  Fig.~\ref{Fig4} and similar ones for quantum critical t$_c$ at fixed h in  a h-t phase diagram. It is shown in the form of t$_c$(h) curves in  Fig.~\ref{Fig5} for two  choices of the g-factors. For (g$_s$,g$_\tau$) = (2,0) and the  cases ($\Delta$,$\delta$)= (0,0), (1,1)  
the scaled t$_c(h)$ is identical close to the zero-field QCP, i.e., the slope does not depend on the anisotropy ($\Delta$,$\delta$). 
 For larger fields one observes the plateau formation in the case ($\Delta$,$\delta$)=(1,1)
and the further monotonic decrease for  ($\Delta$,$\delta$)= (0,0). This behaviour may qualitatively be understood from the field dependence of the local energy levels of $\HH_J$ shown in Fig.~\ref{Fig1}. For  increasing h the gap between the ground state and first excited state decreases
and hence a smaller t$_c$ is necessary to achieve the softening of the triplet excitation at Q. For $\Delta$=1 this effect eventually levels off because the splitting of the two lowest local levels becomes constant at large field and approaches 2$\Delta$J$_x$ so that t$_c(h)$ reaches a plateau at large fields for nonzero $\Delta$. On the other hand for $\Delta$=0 the two lowest levels become asymptotically degenerate and hence t$_c$(h) should approach zero for large fields.  As already noted in the discussion of Fig.~\ref{Fig4} the intermediate case 
 ($\Delta$,$\delta$ ) = (1,0)  and (g$_s$,g$_\tau$) = (2,0) behaves quite differently. After an initial but much weaker decrease of t$_c$(h) it reaches a minimum at (h/J$_x$,t/J$_x$)=(0.275,0.638) and then starts to increase again. The region of the nonmonotonic behaviour with slightly subcritical t/t$_c$(0,1) corresponds to the region where reentrance behaviour KS-AF-KS is observed in Fig.~\ref{Fig4}. For above critical  t/t$_c$(0,1), once t$_c$(h)$>$t the  initial AF phase is suppressed and we obtain the sequence AF-KS, i.e.  a field induced KS phase corresponding to  Fig..~\ref{Fig4} (bottom). In all three cases with (g$_s$,g$_\tau$) = (2,0) one has a finite t$_c$(h) for finite fields. \\

This is different if we chose equal g-factors (g$_s$,g$_\tau$) = (2,2). For this choice the lowest two local levels cross at h=0.25 ($\Delta$=0) and at h=0.5 ($\Delta$=1) (not shown in Fig.~\ref{Fig1}).  This is due to the fact that S$^t_z$ commutes with the Hamiltonian and hence there is no level repulsion of the two S$_z^t$=0 states. Consequently at this point the quantum critical t$_c$(h) vanishes in all three cases of  ($\Delta$,$\delta$). It decreases essentially linearly from its maximum value at h=0. 
The phase diagram in Fig.~\ref{Fig5} shows a collection of the critical t$_c$(h) curves for the six  cases considered. The corresponding gapped KS and AF ordered phases are below and above each curve respectively. We comment again on the influence of third order terms discussed in Sect.~\ref{sec4a} on the phase diagram of Fig.~\ref{Fig5}. Remembering that all cases with $\delta=0$ or $g_s=g_\tau$ are unaffected by the third order contributions, only one of the critical field curves in Fig.~\ref{Fig5} ($\star$ with (g$_s$,g$_\tau$) = (2,0)) may be influenced. In this case third order terms lead to an almost field independent and moderate rescaling of h$_c$ by a factor $\sim$1.2 according to Eq.~(\ref{HCSCALE}) and Fig.~\ref{Fig2}.\\

Finally in Fig.~\ref{Fig6} we show the dependence of the tilting angle of the total moment out of the xy-plane for two possible cases with AF order. It is calculated from Eq.~(\ref{CANTING}) (full line) and a simplified (low field) expression given in the caption. Close to the quantum critical field the staggered moment vanishes and the total moment is then alligned with the external field ($\parallel$ z), meaning $\alpha=\pi/2$.

\section{Discussion}

\label{sec5a}

As we have mentioned in the introduction the interest for the Kondo necklace (KN) model stems from two facts: (i) it is thought to be a simplified version of the original Kondo lattice (KL) model with itinerant electrons and (ii) it is expected to describe some basic features of the observed quantum phase transitions in real Kondo materials. Although there is no mapping or exact equivalence  of the two models the agreement of their qualitative features by simply identifying the intersite interaction (KN) and hopping (KL) energies (both given by t) is generally quite good. As discussed in \cite{Zhang00,Langari06} the 1D Kondo necklace  model  does not exhibit quantum critical behaviour, this agrees with earlier numerical work on the KL model \cite{Scalettar85} where the disordered spin gapped state always prevails \cite{Tsunetsugu97}.\\

We discuss the more interesting 2D case where quantum phase transitions in the KN model are possible \cite{Zhang00,Langari06}. One can compare the values of the quantum critical t$_c$/J (we set J$\equiv$ J$_x$ in this section to comply with literature conventions) obtained for the KN model from Fig.~\ref{Fig3} with those for the KL model obtained in MC simulations obtained previously \cite{Wang94,Assaad99,Capponi01}. In the spirit of Doniach's replacement in 1D we should compare with the KN model for xy-type interactions, i.e. with the $(\delta,\Delta)=(0,1)$ case in Fig.~\ref{Fig3} which has a t$_c$/J=0.7. The MC simulations for the 2D SU(2) fermionic KL model lead to quantum critical parameters  (average values) t$_c$/J= 0.71 \cite{Wang94} , 0.69 \cite{Assaad99,Capponi01} and 0.68 \cite{Beach04}, very close to the appropriate value in Fig.~\ref{Fig3} (upper panel: center of full line; lower panel: left corner of full line) . Note that the critical t$_c$'s corresponding to 
 full lines in Fig.~\ref{Fig3} are obtained from a simple analytical formula within the bond-operator mean field approach \cite{Langari06}. Further support for this scenario comes from exact diagonalisation (ED) results for the KL model on small clusters \cite{Zerec07} using open boundary conditions. The total average local moment $\la\mu_{loc}^2\ra= \la(\boldtau_i+\bS_i)^2\ra$  calculated as function of J/t shows a reduction to half of the free moment size due to Kondo singlet formation in the lattice for the value t$_c$/J = 0.67. If it is interpreted as the critical value in the thermodynamic limit this is again close to the above results.
This near equality of analytical KN and numerical KL results seems to suggest that even in 2D the SU(2) KL model is better described by the U(1) xy-type KN model with ($\delta,\Delta$)=(0,1) since the SU(2) KN model with  ($\delta,\Delta$)=(1,1) has a considerably larger critical value $t_c/J=0.862$. \\
 
 The fermionic KL model is obtained from the periodic Anderson  model (PAM) by a Schrieffer-Wolff transformation which eliminates charge fluctuations of f-electrons. The resulting Kondo coupling J is then given in terms of the Anderson model parameters by $J=-UV/\epsilon_f(\epsilon_f+U)$ where $\epsilon_f<0$ is the f-level position with respect to the Fermi level at zero energy, U is the on-site f-electron repulsion and V their hybridisation strength with conduction electrons. The  periodic Anderson model may be studied with the dynamical mean field theory (DMFT) \cite{Jarrell93}. This method has been used more recently to investigate its quantum critical properties \cite{Sun05} which should be related to those of the Kondo lattice and hence also KN models. The method is formally a D=$\infty$ approximation but has nevertheless been used for studying electron correlations in finite dimensional lattices such as the 3D PAM in \cite{Sun05}. Using the value of J from the Schrieffer-Wolff transformation the extrapolated T= 0 phase diagram suggests that the critical hopping strength at the QCP  is given by $t_c/J\simeq 0.43$. This is again in reasonable agreement with the mean-field bond operator result for the 3D KN model which predicts $t_c/J=0.38$ \cite{Langari06}.\\

While the quantum critical behaviour as function of control parameter t/J is well presented, there are few microscopic investigations concerning the field induced quantum critical behaviour of KL/KN type models such as provided here. The fermionic 2D KL model in an external field with equal g-factors corresponding to (g$_s$,g$_\tau$) = (2,2) was studied in Ref.\onlinecite{Beach04} by using variational and MC methods. The numerical results for J/t =3  suggest that the transition between Kondo and AF ordered regime takes place in a field region (B$\equiv 2h$ in notation of Ref.~\onlinecite{Beach04}) between B$^{-}_c$/t = 1.0 and n B$^{+}_c$/t =2.25. In this regime both Kondo-like features and transverse magnetic order coexist. From the present calculations of the 2D KN model again with ($\delta,\Delta$)=(0,1), g-factors as above and using the data shown in Figs.~\ref{Fig3},\ref{Fig5} one obtains B$_c$/t = 1.86 (B$_c$/J=0.62) which is close to the average 1.63 of
  the above B$_c^{\pm}/t$ values. This value for the critical field was also confirmed by exact diagonalisation (ED) results for the KL model on small clusters \cite{Zerec07} where the breaking up of on-site singlet formation was observed above the value B$_c$/J $\simeq 0.5$. We conclude that  KL and KN models also seem to exhibit similar field induced quantum critical behaviour. Another matter is the scaling behaviour of singlet gap and staggered magnetisation around the critical field. The present mean field type theory does not give sufficient insight into that issue. It has sofar been treated within continuum field theories involving order parameter fluctuations around the critical field \cite{Fischer05}.\\ 
 
The present bond operator mean field results for the KN model are apparently consistent with known numerical results for the fermionic Kondo lattice models. It is less clear whether they are useful for the interpretation of experimental results, especially for the field induced quantum phase transitions. Since charge degrees of freedoms are eliminated in the KN model it is strictly speaking more appropriate for the Kondo-insulator compounds which have a hybridisation gap due to half filled conduction bands. However we ignore this subtlety in the following and also apply it to the magnetism of metallic Kondo compounds.
The zero field quantum critical behaviour as function of the control parameter t/J can experimentally be mimicked by applying hydrostatic or chemical pressure (by substitution of elements). This changes mostly the hybridisation and hence the Kondo coupling J. In the accessible pressure regime one may assume that t/J varies linearly with pressure. In this manner it is feasible to drive AF ordered heavy fermion  systems to the quantum critical point where they become nonmagnetic heavy Fermi liquids. Since hydrostatic (positive) pressure generally increases J it tends to suppress the AF phase while with suitable substitution of elements (negative) chemical pressure may decrease J which favors AF order. Thus the AF QCP may be approached from both sides. There are many examples to be found especially among Ce- compounds, for a review see \cite{Thalmeier05}. In most cases of Ce-compounds, however, the AF QCP may not be reached directly because it is enveloped by a 'dome' of the superconducting phase and the critical pressure has to be obtained from extrapolation to T=0. While the KL/KN type models discussed above naturally suggest the AF QCP there is no direct way of experimental determination of the critical (t/J)$_c$ since its relation to the experimentally accessible critical pressure p$_c$ (or critical concentration of substituent) is unknown. In fact most of the interest on pressure induced QCP's focuses more on the scaling exponents of various quantities with respect to the distance (p-p$_c$) to the QCP. This is extensively discussed in the reviews cited in the introduction.\\

For the field induced quantum phase transition much less experimental results are available. A review of materials investigated is given in Ref. \onlinecite{Stewart01}.
A recent example of a field induced destruction of AF order is the tetragonal YbRh$_2$Si$_2$ compound \cite{Gegenwart03}. At ambient pressure and zero field it has an AF order which is indeed of the easy plane (xy)-interaction type ($\delta$=0). The AF order is destroyed at 0.66T for field along the hard c-axis and a nonmagnetic Fermi liquid state appears above the critical field . If one assumes that the local Kondo interaction which results from the Schrieffer-Wolff mechanism is more of the isotropic nature ($\Delta$=1) it qualitatively corresponds to the AF-KS scenario shown in the lower panel of Fig.~\ref{Fig4}. As mentioned before the reentrant scenario shown also in this figure exists only in a narrow parameter range and it is perhaps understandable that no such example is known. What is really surprising is the following observation: Most of the parameter cases in Fig.~\ref{Fig5} would predict the KS-AF sequence, i.e. the field-induced AF order out of the Kondo phase. This is a natural consequence that in most cases a field reduces the singlet-triplet gap and supports the onset of magnetic order as shown in the upper panel of Fig.~\ref{Fig4}. We would like to stress that this is also the only phase sequence found in the fermionic KL model \cite{Beach04}.
 However field induced AF magnetic order is not easily found in heay fermion compounds. For example CeNi$_2$Ge$_2$ at ambient pressure and zero field  is in the Kondo singlet phase and rather close to the AF quantum phase transition which may be achieved by appropriate substitution of Ni. Thus it is a complementary case to  YbRh$_2$Si$_2$.
However in an external field there is no field induced transition to transverse AF order as one might naively expect from the model discussion, rather it is driven further away from the AF QCP as specific heat and resistivity measurements suggest \cite{Gegenwart03}. There is however one known example of a heavy fermion system which exhibits field induced AF order, though complicated by the appearance of superconductivity. The compound CeRhIn$_5$  is already an AF at ambient pressure \cite{Park06}. Application of hydrostatic pressure  moves it to the nonmagnetic (KS) side where it also becomes superconducting. Additional application of a magnetic field destroys the superconductivity and leads to the reestablishment of a field induced AF order in a wide range of pressure form 1.4 to 2.4 GPa \cite{Park06}. This is indeed the KS-AF sequence of phases corresponding to Fig.~\ref{Fig4} (top) which appears in the majority of cases studied here. Experimentally, however, it seems to be the one most rarely encountered.\\

These examples suggest that while the Kondo lattice or necklace type models have some qualitative and instructive properties which are relevant for magnetic quantum phase transitions they may not be able to explain some experimental observations in heavy fermion compounds. It is well possible that the complete elimination of charge degrees of freedom is too radical  by suppressing hybridisation effects of moments with conduction electrons and thus favoring the magnetic phases. Therefore the more fundamental PAM model, including appropriate local and conduction electron Zeeman terms may be a more promising starting point for future investigations of field induced quantum phase transitions in heavy fermion compounds.

\section{Conclusion and Outlook}
\label{sec6}
In this work we have investigated the general anisotropic Kondo necklace model and its quantum phase transition from 
paramagnetic singlet phase to AF phase and vice versa  in an external field. We have derived a bosonic Hamiltonian using the bond operator representation of local and interacting spins and diagonalised the bilinear part by a generalised Bogoliubov transformation to circular polarised  modes to include the effect of the external field. Higher order parts of the Hamiltonian were found to be insignificant for the quantum critical behaviour. In mean field approximation one obtains a total ground state energy that depends on the singlet amplitude $\bs$, staggered and uniform triplet amplitudes $\bt,\bth$ and the chemical potential. Minimization leads to two sets of selfconsistent equations for Kondo singlet and AF phase, respectively.

 From their numerical solution we find that in most investigated cases the t$_c$ value of  the quantum critical point decreases monotonously with field strength leading to the KN-AF phase sequence for increasing field. In the case of unequal g-factors and  ($\Delta$,$\delta$) = (1,1) it levels off at a plateau value t$_c(h)$/t$_c(\Delta,\delta)\simeq$ 0.4 while for  ($\Delta$,$\delta$) = (0,0) it continues to decrease at larger field. However for the genuine KN case with ($\Delta$,$\delta$) = (1,0) 
the t$_c$(h) curve is nonmonotonic with a minimum at $h/J_x\simeq$ 0.275 at small fields and continues to increase for larger fields. Depending on the size of t this implies two possibilities:
 For above critical t a suppression of AF and the opening of a spin excitation gap for $h> h_c$ (AF-KN sequence) or the reentrance behaviour (KN-AF-KN sequence) of the gapped singlet phase for slightly subcritical t (as compared to t$_c$(0)). For equal g-factors a crossing of local levels occurs due to a conserved S$_z^t$ and the critical t$_c$ reaches zero already at relatively small fields for all investigated cases of ($\delta,\Delta$). Since our treatment is of mean-field type we will get mean-field exponents for the spin gap E$_g$ and magnetisation m$_s$ at the field-induced QCP. This is apparent from Fig.\ref{Fig4} and has not been discussed further. Improvement for the critical exponents will require a selfconsistent renormalisation theory for the triplet  excitations close to the quantum critical point.

\section*{Acknowledgment}
The authors would like to thank V. Yushankhai for helpful suggestions. A. L. would like to acknowledge the
support of Sharif University of Technology.  A. L.  also would like to acknowledge the Max Planck Institute for the
Physics of Complex Systems - Dresden where the initial part of this work was started during his visit.



\end{document}